\documentclass[12pt]{article}
\usepackage{a4}
\usepackage{amsmath}
\usepackage{amssymb}
\usepackage{latexsym}
\usepackage{amssymb}
\usepackage{epsfig}
\usepackage{natbib}
\def \pd{\partial}

\def \e{{\mathrm{e}}}

\def \tl#1{\overset{\kern 1pt\circ}{#1}}
\def \TL#1{\overset{\kern -3pt \circ}{#1}}
\def \TLL#1{\overset{\kern -7pt \circ}{#1}}

\def \Bphi{\boldsymbol{\phi}}

\def \Bphi{{\boldsymbol{\phi}}}

\def \Ba{{\boldsymbol{a}}}

\def \Bu{{\boldsymbol{u}}}

\def \Bx{{\boldsymbol{x}}}

\def \LL{{\cal{L}}}

\begin{document}
\title{{\bf A screw dislocation in a functionally graded material using the 
translation gauge theory of dislocations}}
\author{
Markus Lazar~$^\text{a,b,}$\footnote{
{\it E-mail address:} lazar@fkp.tu-darmstadt.de (M.~Lazar).} 
\\ \\
${}^\text{a}$ 
        Heisenberg Research Group,\\
        Department of Physics,\\
        Darmstadt University of Technology,\\
        Hochschulstr. 6,\\      
        D-64289 Darmstadt, Germany\\
${}^\text{b}$ 
Department of Physics,\\
Michigan Technological University,\\
Houghton, MI 49931, USA
}

\date{\today}    
\maketitle


\begin{abstract}
The aim of this paper is to provide new results and insights for a screw 
dislocation in functionally graded media within the gauge theory of 
dislocations. 
We present the equations of motion for dislocations in inhomogeneous media.
We specify the equations of motion for a screw dislocation in a
functionally graded material. The material properties are assumed to vary 
exponentially along the $x$ and $y$-directions. 
In the present work we give the analytical gauge field theoretic solution 
to the problem of a screw dislocation in inhomogeneous media.
Using the dislocation gauge approach, 
rigorous analytical expressions for the elastic distortions, 
the force stresses, the dislocation density and the pseudomoment stresses 
are obtained depending on the moduli of gradation and an effective 
intrinsic length scale characteristic for the functionally graded material
under consideration.
\\

\end{abstract}

\section{Introduction}

Nonhomogeneous media, multilayered structures and 
functionally graded materials (FGMs)
are of considerable technical and engineering importance as well as of mathematical interest (see, e.g., 
\citet{E95}). 
Generally, such materials 
refer to heterogenous composite materials in which 
the material moduli vary smoothly and continously from point to point.
Typical examples of FGMs are ceramic/ceramic and metal/ceramic systems.
Compressive reviews on aspects of FGMs can be found in \citet{M95} 
and \citet{E95}.
However, only few investigations have been made to assess the role and 
the importance of dislocations in FGMs.
For the first time, 
\citet{Barnett} found the displacement and stress fields of a screw dislocation
in an isotropic medium which is arbitrary graded in the $x$-direction. 
One important example for a FGM are exponentially graded materials. For such
media the elasticity moduli are exponential functions depending on
the space coordinates and the new material parameters describing the gradation
(see, e.g., \citet{E95}).
The Green function for a two dimensional exponentially graded elastic medium
has been found by~\citet{Paulino04} and the 
three-dimensional Green function is given by~\citet{Martin02}. 

Classical continuum theories possess no intrinsic length scales and, therefore,
they are scale-free continuum theories 
which are not able to describe size effects being of importance 
at micro- and nanoscales and near defects.
In order to describe such effects generalized continuum theories are needed.
Such generalized continuum theories are, for instance, strain gradient 
theories~\citep{Mindlin64,AA97,LM05}
and the dislocation gauge theory~\citep{Edelen83,Edelen88,Lazar02,LA09} 
which enrich the classical theories with additional material lengths.
Using strain gradient elasticity, cracks in FGMs have been investigated 
by~\citet{Paulino03} and \citet{Chan08}.
For the first time ever, the solution of a screw dislocation in FGMs, using 
the theory of gradient elasticity, was given by~\citet{Lazar07}.
Gradient elasticity is a theory which is very popular in engineering sciences.
But not everything is well-understood about the physical importance
of the higher-order stresses (hyperstresses).
Gradient elasticity should be understood as an effective theory since
it was invented for compatible deformations.
On the other hand, a more physically motivated `gradient-like' theory
is the so-called dislocation gauge theory.
In the dislocation gauge theory both the incompatible elastic distortion tensor
and the dislocation density tensor are physical state quantities
giving contribution to the elastic stored energy density. 
Also for nonhomogeneous media these physical state quantities are 
gauge-invariant. 
In this approach, the higher-order stress is realized as pseudomoment stress
which is the response to the presence of dislocations. This pseudomoment 
stress is related to the moment stress known from generalized elasticities
like Cosserat elasticity~(see, e.g., \citet{Nowacki,Eringen}).
Thus, the force stress tensor is asymmetric like in Cosserat elasticity.
In general, in gradient elasticity the hyperstress (double stress) possesses
five material moduli~\citep{Mindlin64}. 
In a simplified version of gradient elasticity~\citep{AA97}
under ad-hoc assumptions the double stress can be simplified.
In the dislocation gauge theory such assumptions are not necessary.
For that reason we use the gauge theory of dislocations in this paper.

In this paper we study a screw dislocation in FGM 
in the framework of the translation gauge theory of dislocations.
In Section 2 we start with the general framework of the translational gauge theory
of dislocations and we derive the isotropic three-dimensional 
field equations of dislocations
for nonhomogeneous media as well as for exponentially graded materials.
In Section 3 
we specialize in the anti-plane problem and 
we give the gauge-theoretical solution of a
screw dislocation  in exponentially graded materials. 
We find rigorous and `simple' analytical solutions
for the force stresses, elastic distortion, torsion and pseudomoment stresses 
produced by a screw dislocation in FGMs. 
Our hope is that the gauge solution will provide new physical insight to possible
improvements of designing FGMs.

\section{Gauge theory of dislocations}
This section introduces the notation and constitutive equations 
of the translational gauge theory of dislocations~\citep{LA09}, 
which will be used to
investigate a screw dislocation in FGMs.
In the dislocation gauge theory, the elastic distortion tensor $\beta_{ij}$ 
is incompatible due to the occurrence of a so-called translational gauge
field $\phi_{ij}$
\begin{align}
\label{beta}
\beta_{ij}=u_{i,j}+\phi_{ij}\, , 
\end{align}
where $u_i$ is the displacement vector. 
It is important to note that $u_i$ and $\phi_{ij}$ are just the
canonical field quantities and they are not unique and not 
physical state quantities.
In presence of dislocations, $u_i$ and $\phi_{ij}$
are discontinuous (or multivalued) fields.
The gauge field $\phi_{ij}$ may be identified with the negative plastic distortion.
The appearance of incompatibility in form of the translation gauge field
gives rise to an incompatibility tensor in the framework of the 
translation gauge theory of dislocations which is called the dislocation 
density tensor or torsion tensor.
Accordingly, the dislocation density tensor is given in terms of the gauge field
\begin{align}
\label{T1}
T_{ijk}=\phi_{ik,j}-\phi_{ij,k}\,, \qquad T_{ijk}=-T_{ikj}
\end{align}
and, alternatively, in terms of the elastic distortion 
\begin{align}
\label{T2}
T_{ijk}
=\beta_{ik,j}-\beta_{ij,k}\, .
\end{align}
In the present framework, the elastic distortion tensor $\beta_{ij}$ and
the dislocation density tensor $T_{ijk}$ are the physical state quantities.
Due to the form of the dislocation density tensor (\ref{T1}) and (\ref{T2})
it fulfills the translational Bianchi identity
\begin{align}
\label{BI}
\epsilon_{jkl} T_{ijk,l}=0\, .
\end{align}
Eq.~(\ref{BI}) is the well-known conservation law of dislocations 
and states that a dislocation line cannot end inside the body.

In the static case
of the (linear) translation gauge theory of dislocations, the Lagrangian
density is of the bilinear form 
\begin{align}
\label{tot-Lag}
\LL=-W=
-\frac{1}{2}\,\sigma_{ij}\beta_{ij}  
- \frac{1}{4}\,H_{ijk}T_{ijk}\,,
\end{align}
where $W$ denotes the stored energy density.
The canonical conjugate quantities (response quantities) are defined by
\begin{align}
\label{can-qua}
\sigma_{ij}:=-\frac{\pd \LL}{\pd \beta_{ij}}\, ,\qquad
H_{ijk}:=-2\frac{\pd \LL}{\pd T_{ijk}}\,,\qquad H_{ijk}=-H_{ikj}\,,
\end{align}
where $\sigma_{ij}$ and $H_{ijk}$ are 
the force stress tensor  and the pseudomoment stress tensor, respectively.
Here, the force stress tensor $\sigma_{ij}$ is asymmetric.
The moment stress tensor $\tau_{ijk}=-\tau_{jik}$ 
can be obtained from the 
pseudomoment stress tensor: $\tau_{ijk}=-H_{[ij]k}$ (see \citet{LA09,LH}).
They have the dimensions:
$[\sigma_{ij}]=\text{force}/(\text{length})^2
\stackrel{\rm SI}{=}\text{Pa}$ and 
$[H_{ijk}]=\text{force}/\text{length}\stackrel{\rm SI}{=}\text{N}/\text{m}$.

The Euler-Lagrange equations with respect to the canonical field variables
derived from the total Lagrangian density
$\LL=\LL(\beta_{ij},T_{ijk})$ are given by
\begin{align}
\label{EL-1}
&E^{\,\Bu}_i (\LL)=  \pd_j \frac{\pd \LL}{\pd u_{i,j}}  
- \frac{\pd \LL}{\pd u_{i}} = 0\, ,\\
\label{EL-2}
&E^{\, \Bphi}_{ij}(\LL)=
 \pd_k \frac{\pd \LL}{\pd \phi_{ij,k}} - \frac{\pd \LL}{\pd \phi_{ij}}  = 0\,.
\end{align}
We may add to $\LL$ a so-called null Lagrangian, $\LL_N=\sigma^0_{ij}\beta_{ij}$, 
with the `background' stress $\sigma^0_{ij}$ satisfying: $\sigma^0_{ij,j}= 0$.
In terms of the canonical conjugate quantities~(\ref{can-qua}),
Eqs.~(\ref{EL-1}) and (\ref{EL-2}) take the form
\begin{alignat}{2}
\label{EL-1b}
\sigma_{ij,j}&= 0\,,
\qquad &&(\text{force balance of elasticity})\, ,\\
\label{EL-2b}
 H_{ijk,k}+ \sigma_{ij}&=\sigma^0_{ij}\,,
\qquad &&(\text{stress balance of dislocations})\,.
\end{alignat}
The force stresses are the sources of the pseudomoment stress.
In fact, one can see in Eq.~(\ref{EL-2b}) that 
the source of the pseudomoment stress tensor is an effective force stress tensor
$(\sigma_{ij}-\sigma^0_{ij})$ (see, e.g., \citet{Edelen88}). 
This effective stress tensor, 
which is the difference between the gauge-theoretical stress field and the 
background stress field, drives the dislocation fields.

The linear, 
isotropic constitutive relations for the force stress and
the pseudomoment stress are
\begin{align}
\label{CE-1}
\sigma_{ij}&= \lambda \delta_{ij} \beta_{kk} + \mu (\beta_{ij}+\beta_{ji}) 
+ \gamma  (\beta_{ij}-\beta_{ji})\,,\\
\label{CE-2}
H_{ijk}&= c_1 T_{ijk} + c_2  (T_{jki} - T_{kji}) + c_3 (\delta_{ij}T_{llk} - \delta_{ik}T_{llj})\, .
\end{align}
Here $\mu, \lambda, \gamma$ are the elastic stiffness moduli and
$c_1,c_2,c_3$ denote the resistivity moduli associated with dislocations.
The six material moduli have the dimensions:
$[\mu, \lambda, \gamma]=\text{force}/(\text{length})^2
\stackrel{\rm SI}{=}\text{Pa}$ and 
$[c_1,c_2,c_3]=\text{force}\stackrel{\rm SI}{=}\text{N}$.
Here, the material moduli are nonhomogeneous that means they depend on the 
space coordinates: $\mu\equiv \mu(\Bx)$, $\lambda\equiv \lambda(\Bx)$, 
$\gamma\equiv \gamma(\Bx)$, $c_1\equiv c_1(\Bx)$, $c_2\equiv c_2(\Bx)$ and 
$c_3\equiv c_3(\Bx)$.
The constitutive relations for nonhomogeneous media~(\ref{CE-1}) and 
(\ref{CE-2}) have the same formal form as the relations for homogeneous media. 
The only difference is that the constitutive moduli are not constant.

The requirement of non-negativity of the stored 
energy (material stability) $W\ge~0$ 
leads to the conditions of semi-positiveness of the constitutive moduli.
Particularly, the constitutive moduli have to fulfill 
the following conditions~\citep{LA09}
\begin{alignat}{3}
\label{IE-mu}
\mu&\ge0\,,\qquad &\gamma&\ge 0\,,\qquad &3\lambda+2\mu&\ge 0\,,\\
\label{IE-c}
c_1-c_2&\ge 0\, ,\qquad &c_1+2c_2&\ge 0\, , \qquad &c_1-c_2+2 c_3&\ge 0\,.
\end{alignat}

If we substitute the constitutive equations~(\ref{CE-1}) and (\ref{CE-2}) 
into the Euler-Lagrange equations~(\ref{EL-1b}) and (\ref{EL-2b}) 
and use the definition~(\ref{T2}), we obtain the explicite form
of the Euler-Lagrange equations for a nonhomogeneous material
\begin{align}
\label{EOM-1}
& \lambda \beta_{jj,i}
+(\mu+\gamma)\beta_{ij,j}+ (\mu-\gamma)\beta_{ji,j}
+\lambda_{,i} \beta_{jj}
+(\mu+\gamma)_{,j}\beta_{ij}+ (\mu-\gamma)_{,j}\beta_{ji}=0\, ,\\ 
\label{EOM-2}
& c_1(\beta_{ik,jk}-\beta_{ij,kk}) 
+c_2(\beta_{ji,kk}-\beta_{jk,ik}+\beta_{kj,ik}-\beta_{ki,jk}) 
+c_3\big[\delta_{ij}(\beta_{lk,lk}-\beta_{ll,kk})
+ \beta_{kk,ji}-\beta_{kj,ki}\big] \nonumber\\
&\ 
+c_{1,k}(\beta_{ik,j}-\beta_{ij,k})
+c_{2,k}(\beta_{ji,k}-\beta_{jk,i}+\beta_{kj,i}-\beta_{ki,j})
+c_{3,k}\, \delta_{ij}(\beta_{lk,l}-\beta_{ll,k})
+c_{3,i} (\beta_{kk,j}-\beta_{kj,k})
\nonumber\\
&\ +\lambda\, \delta_{ij}(\beta_{kk}-\beta^0_{kk}) 
+(\mu+\gamma)(\beta_{ij}-\beta^0_{ij})
+(\mu-\gamma)(\beta_{ji}-\beta^0_{ji})=0\, ,
\end{align} 
where $\beta_{ij}^0$ denotes the `classical' elastic distortion tensor.
It can be seen that 
the space-dependence of the material moduli influences the field equations~(\ref{EOM-1}) and (\ref{EOM-2}) due to gradients of the material moduli.
Eq.~(\ref{EOM-1}) is a partial differential equation of first order
for the elastic distortion as well as for the constitutive moduli.
On the other hand, Eq.~(\ref{EOM-2}) contains differential 
operators of second order acting on the elastic distortion and
`mixed' gradients of the constitutive moduli and the elastic distortions.

Rather than looking for the general inhomogeneous behaviour of the
constitutive moduli, we specify here on an inhomogeneous material,
which can find application to functionally graded materials. 
We assume that the material properties very in a simple, explicite manner.
Here, we consider exponential variations of the six constitutive moduli
in the following manner
\begin{align}
\label{FGM-cons-3d}
&\lambda=\lambda^0\, \e^{2(a_1 x+a_2 y+a_3 z)}\, ,\quad
\mu=\mu^0\, \e^{2(a_1 x+a_2 y+a_3 z)}\, ,\quad
\gamma=\gamma^0\, \e^{2(a_1 x+a_2 y+a_3 z)}\, ,\nonumber\\
&c_1=c_1^0\, \e^{2(a_1 x+a_2 y+a_3 z)}\, ,\quad
c_2=c_2^0\, \e^{2(a_1 x+a_2 y+a_3 z)}\, , \quad
c_3=c_3^0\, \e^{2(a_1 x+a_2 y+a_3 z)}
\, ,
\end{align}
where $\lambda ^0$, $\mu^0$, $\gamma^0$, $c_1^0$, $c^0_2$, $c^0_3$, 
$a_1$, $a_2$ and $a_3$ are the material parameters of the FGM. 
The moduli $a_1$, $a_2$ and $a_3$ are the characteristic parameters for 
exponentially gradation in $x$, $y$ and $z$-direction
and they have the dimensions:
$[a_1,a_2,a_3]=1/\text{length}$.
The parameters $a_1$, $a_2$ and $a_3$ determine 
the gradation of all six constitutive moduli $\lambda$, $\mu$, $\gamma$, $c_1$, $c_2$, $c_3$.
For such an exponentially FGM, the equations (\ref{EOM-1}) and (\ref{EOM-2}) 
reduce to
\begin{align}
\label{EOM-3}
& \lambda (\nabla_i+2a_i)\beta_{jj}
+(\mu+\gamma) (\nabla_j+2a_j)\beta_{ij}
+ (\mu-\gamma) (\nabla_j+2a_j)\beta_{ji}=0\, ,\\ 
\label{EOM-4}
& c_1 (\nabla_k+2a_k) (\nabla_j \beta_{ik}-\nabla_k\beta_{ij}) 
+c_2  (\nabla_k+2a_k)(\nabla_k \beta_{ji}-\nabla_i \beta_{jk}
+\nabla_i\beta_{kj}-\nabla _j\beta_{ki}) \nonumber\\
&\ +c_3\, \delta_{ij} (\nabla_k+2a_k)(\nabla_l\beta_{lk}-\nabla_k\beta_{ll})
+c_3  (\nabla_i+2a_i) (\nabla_j \beta_{kk}-\nabla_k\beta_{kj}) \nonumber\\
&\ +\lambda\, \delta_{ij}(\beta_{kk}-\beta^0_{kk}) 
+(\mu+\gamma)(\beta_{ij}-\beta^0_{ij})
+(\mu-\gamma)(\beta_{ji}-\beta^0_{ji})=0\, ,
\end{align} 
where $\nabla_i$ denotes the three-dimensional Nabla operator.
Eqs. (\ref{EOM-3}) and (\ref{EOM-4}) are the three-dimensional governing equations of
dislocations in exponentially graded materials using the dislocation gauge theory.

\section{Gauge solution of a screw dislocation in functionally graded materials}

We now derive the equations of motion
for a screw dislocation in FGMs. We consider
an infinitely long screw dislocation parallel to the $z$-axis with
the Burgers vector $b=b_z$.
The symmetry of such a straight screw dislocation 
leaves only the following non-vanishing 
components of the elastic distortion tensor (see,
e.g., \citet{deWit}):
$\beta_{zx}$, $\beta_{zy}$, and for the dislocation density tensor: 
$T_{zxy}$. 

For the anti-plane problem of a screw dislocation in nonhomogeneous materials, 
the equilibrium condition~(\ref{EOM-1}) reduces to
\begin{align}
\label{EC-FGM}
\beta_{zj,j}=-\frac{(\mu+\gamma)_{,j}}{\mu+\gamma}\, \beta_{zj}\,,
\qquad j=x,y\, .
\end{align}
Using equation~(\ref{EC-FGM}), we obtain from (\ref{EOM-2})  the
following equations for the components $\beta_{zx}$ and $\beta_{zy}$
of a screw dislocation in nonhomogeneous materials
\begin{align}
\label{EOM-Bzx1}
c_1 \beta_{zx,jj}+c_{1,j}\beta_{zx,j}
-\Big(c_{1,j}-c_1\, \frac{(\mu+\gamma)_{,j}}{\mu+\gamma}\Big)\beta_{zj,x}
-(\mu+\gamma)(\beta_{zx}-\beta^0_{zx})&=0\, ,\\
\label{EOM-Bzx2}
c_2 \beta_{zx,jj}+c_{2,j}\beta_{zx,j}
-\Big(c_{2,j}-c_2\, \frac{(\mu+\gamma)_{,j}}{\mu+\gamma}\Big)\beta_{zj,x}
+(\mu-\gamma)(\beta_{zx}-\beta^0_{zx})&=0\, ,\\
\label{EOM-Bzy1}
c_1 \beta_{zy,jj}+c_{1,j}\beta_{zy,j}
-\Big(c_{1,j}-c_1\, \frac{(\mu+\gamma)_{,j}}{\mu+\gamma}\Big)\beta_{zj,y}
-(\mu+\gamma)(\beta_{zy}-\beta^0_{zy})&=0\, ,\\
\label{EOM-Bzy2}
c_2 \beta_{zy,jj}+c_{2,j}\beta_{zy,j}
-\Big(c_{2,j}-c_2\, \frac{(\mu+\gamma)_{,j}}{\mu+\gamma}\Big)\beta_{zj,y}
+(\mu-\gamma)(\beta_{zy}-\beta^0_{zy})&=0\, .
\end{align}
For the two-dimensional anti-plane problem of FGMs, 
we assume that the constitutive moduli are exponentially graded,
\begin{align}
\label{FGM-cons}
\mu=\mu^0\, \e^{2(a_1 x+a_2 y)}\, ,\quad
\gamma=\gamma^0\, \e^{2(a_1 x+a_2 y)}\, ,\quad
c_1=c_1^0\, \e^{2(a_1 x+a_2 y)}\, ,\quad
c_2=c_2^0\, \e^{2(a_1 x+a_2 y)}\, ,
\end{align}
where $\mu^0$, $\gamma^0$, $c_1^0$, $c^0_2$, $a_1$ and $a_2$ are the material 
constants of the anti-plane problem. 
The moduli $a_1$ and $a_2$ are the characteristic parameters of exponentially gradation in $x$ and $y$-direction, respectively.
If we substitute (\ref{FGM-cons}) into (\ref{EOM-Bzx1})--(\ref{EOM-Bzy2}),
the governing equations are
\begin{align}
\label{EOM-Bzx-fgm-1}
c_1(\Delta+2 \Ba\cdot\!\nabla)\beta_{zx}-(\mu+\gamma)(\beta_{zx}-\beta^0_{zx})&=0\,, \\
\label{EOM-Bzx-fgm-2}
c_2(\Delta+2 \Ba\cdot\!\nabla)\beta_{zx}+(\mu-\gamma)(\beta_{zx}-\beta^0_{zx})&=0\,, \\
\label{EOM-Bzy-fgm-1}
c_1(\Delta+2 \Ba\cdot\!\nabla)\beta_{zy}-(\mu+\gamma)(\beta_{zy}-\beta^0_{zy})&=0\,, \\
\label{EOM-Bzy-fgm-2}
c_2(\Delta+2 \Ba\cdot\!\nabla)\beta_{zy}+(\mu-\gamma)(\beta_{zy}-\beta^0_{zy})&=0\, ,
\end{align}
where $\Delta$ and $\nabla$ denote the two-dimensional 
Laplacian and the Nabla operator, respectively.
It can be seen in (\ref{EOM-Bzx-fgm-1})--(\ref{EOM-Bzy-fgm-2}) that we have four
equations for only two components $\beta_{zx}$ and $\beta_{zy}$.
Thus, it is clear that not all four equations (\ref{EOM-Bzx-fgm-1})--(\ref{EOM-Bzy-fgm-2})
are independent. In fact, if we compare (\ref{EOM-Bzx-fgm-1}) with (\ref{EOM-Bzx-fgm-2})
as well as (\ref{EOM-Bzy-fgm-1}) with (\ref{EOM-Bzy-fgm-2}), we find the 
relation between the material moduli
\begin{align}
\label{Rel-c}
\frac{c_1}{\mu+\gamma}=-\frac{c_2}{\mu-\gamma}\, .
\end{align}
In addition, we may define the following intrinsic gauge length of 
the anti-plane problem of a screw dislocation 
\begin{align}
\label{L}
\ell=\sqrt{\frac{c_1}{\mu+\gamma}}
=\sqrt{\frac{c^0_1}{(\mu+\gamma)^0}}
\, .
\end{align}
Evidently, the length scale $\ell$ is constant for FGMs using (\ref{FGM-cons}).
This fact is in agreement with gradient elasticity for 
FGMs~\citep{Paulino03,Chan06,Chan08,Lazar07} where the characteristic 
gradient length is also constant. Only the constitutive moduli entering the constitutive relations are nonconstant.

By means of~(\ref{Rel-c}) and (\ref{L}) the material moduli 
$c_1$ and $c_2$ can be expressed 
in terms of the length $\ell$, and the constitutive moduli $\mu$ and $\gamma$
\begin{align}
c_1=\ell^2 (\mu+\gamma)\, ,\qquad
c_2=-\ell^2 (\mu-\gamma)\, .
\end{align}
Thus, the gradation of $c_1$ and $c_2$ is given by the gradation of the 
constitutive moduli $\mu$ and $\gamma$.
Accordingly, the governing equations for $\beta_{zx}$ and $\beta_{zy}$ reduce to
\begin{align}
\label{Bzx-pde}
[1-\ell^2(\Delta+2 \Ba\cdot\!\nabla)]\beta_{zx}&=\beta^0_{zx}\,,\\
\label{Bzy-pde}
[1-\ell^2(\Delta+2 \Ba\cdot\!\nabla)]\beta_{zy}&=\beta^0_{zy}\, .
\end{align}
Due to the gradation term $\Ba\cdot\!\nabla$, Eqs. (\ref{Bzx-pde}) and (\ref{Bzy-pde})
may be called `perturbed' Helmholtz equations.

Differentiating equation~(\ref{Bzx-pde}) with respect to $y$ and 
equation~(\ref{Bzy-pde}) with respect to $x$ and using the definition of the 
torsion (\ref{T2}),
the governing equation for the torsion (dislocation density) of a screw 
dislocation in FGM turns out to be 
\begin{align}
\label{T-pde}
[1-\ell^2(\Delta+2 \Ba\cdot\!\nabla)]\,T_{zxy}&=T^0_{zxy}\,,
\end{align}
where 
\begin{align}
\label{T0}
T^0_{zxy}=b\, \delta(x)\delta(y)
\end{align}
is the dislocation density of a straight screw dislocation in classical 
elasticity.
Using the substitution
\begin{align}
T_{zxy}=\e^{-(a_1 x+a_2y)}\, \Psi\, ,
\end{align}
we obtain from Eq. (\ref{T-pde}) the following Helmholtz equation
\begin{align}
\label{Psi-pde}
[\Delta-\kappa^2]\,\Psi&= -\frac{b}{\ell^2}\, \delta(x)\delta(y)\,,
\end{align}
with
\begin{align}
\label{kappa}
\kappa=\sqrt{\frac{1+a^2\ell^2}{\ell^2}}\, ,\qquad
a=\sqrt{a_1^2+a_2^2}\, .
\end{align}
It can be seen that $\kappa$ is the inverse intrinsic length 
characteristic for the anti-plane problem of 
FGMs in the dislocation gauge approach.
Therefore, $1/\kappa$ is an effective length scale given in terms of the 
gauge length scale $\ell$ and the gradation length scale $1/a$.
The inverse length scale of gradation $a$ is also constant.
Thus, in addition to $1/a$, two (constant) characteristic length scales, 
namely $\ell$ and $1/\kappa$,
appear in the gauge-theoretical approach of the anti-plane problem of FGMs. 
The solution of (\ref{Psi-pde}) is given by
\begin{align}
\Psi=\frac{b}{2\pi\ell^2}\, K_0(\kappa r)\, ,
\end{align}
where $r=\sqrt{x^2+y^2}$ 
and $K_n$ is the $n$th order modified Bessel function of the second kind. 
Finally, the torsion of a screw dislocation in FGM using the dislocation gauge
theory is obtained as
\begin{align}
\label{T-screw}
T_{zxy}=\frac{b}{2\pi\ell^2}\, \e^{-(a_1 x+a_2 y)}\, K_0(\kappa r)\, .
\end{align}
If we compare the dislocation density~(\ref{T-screw}) with the corresponding 
one for
homogeneous media (see, e.g., \citet{LA09}), we observe the exponential factor
and the length scale $\kappa$ instead of $1/\ell$ as difference.
That means the dislocation density~(\ref{T-screw}) has lost the cylindrical symmetry and decays 
faster in the far-field in FGMs.
Substituting (\ref{T-screw}) into the constitutive relation~(\ref{CE-2}) 
and making use of (\ref{Rel-c}), we obtain for the components of the
pseudomoment stress tensor
\begin{align}
\label{H-zxy}
H_{zxy}&=\frac{(\mu+\gamma)^0\, b}{2\pi}\, \e^{(a_1 x+a_2 y)}\, K_0(\kappa r)\,, \\
\label{H-xyz}
H_{xyz}&=-\frac{(\mu-\gamma)^0\, b}{2\pi}\, \e^{(a_1 x+a_2 y)}\, K_0(\kappa r)\,, \\
\label{H-yzx}
H_{yzx}&=-\frac{(\mu-\gamma)^0\, b}{2\pi}\, \e^{(a_1 x+a_2 y)}\, K_0(\kappa r)\,. 
\end{align}
Due to the gradation, the fields (\ref{T-screw})--(\ref{H-yzx})
do not possess cylindrical symmetry in contrast to the results
for homogeneous media.
It can be seen 
that the expressions (\ref{T-screw})--(\ref{H-yzx}) for 
$a_1\rightarrow 0$ and $a_2\rightarrow 0$ coincide with the result obtained by
\citet{LA09} for homogeneous materials.

To solve the remaining equations~(\ref{Bzx-pde}) and (\ref{Bzy-pde}), 
we use the constitutive relations~(\ref{CE-1}) for the components
$\sigma_{zx}$ and $\sigma_{zy}$. In order to fulfill the 
equilibrium condition for the forces stresses $\sigma_{zx,x}+\sigma_{zy,y}=0$,
we introduce the stress function $F$ according to
\begin{align}
\label{sigma-zx-F}
\sigma_{zx}&=(\mu+\gamma)\beta_{zx}=-F_{,y}\, ,\\
\label{sigma-zy-F}
\sigma_{zy}&=(\mu+\gamma)\beta_{zy}=F_{,x}\, .
\end{align}
By the help of equations~(\ref{sigma-zx-F})
and (\ref{sigma-zy-F}), we are able to express the elastic
distortion $\beta_{zx}$ and $\beta_{zy}$ in terms of the stress function $F$
according
\begin{align}
\label{Bzx-F}
\beta_{zx}&=-\frac{1}{\mu+\gamma}\, F_{,y}\, ,\\
\label{Bzy-F}
\beta_{zy}&=\frac{1}{\mu+\gamma}\, F_{,x}\, .
\end{align}
Of course, Eqs.~(\ref{Bzx-F}) and (\ref{Bzy-F}) satisfy the 
condition~(\ref{EC-FGM}). 
If we substitute (\ref{Bzx-F}) and (\ref{Bzy-F})
into (\ref{Bzx-pde}) and (\ref{Bzy-pde}), we find the following partial differential equation for the stress function $F$
\begin{align}
\label{F-pde}
[1-\ell^2(\Delta-2 \Ba\cdot\!\nabla)] F&=F^0\,,
\end{align}
where $F^0$ on the right hand side satisfies the equation
\begin{align}
\label{F0-pde}
[\Delta-2 \Ba\cdot\!\nabla]F^0=(\mu+\gamma)^0\, b\, \delta(x)\delta(y)\, .
\end{align}
Eq.~(\ref{F0-pde}) is a `perturbed' Laplace equation.
The solution of (\ref{F0-pde}), which is actually the Green function,
reads (see \citet{Lazar07,WP08})
\begin{align}
F^0=-\frac{(\mu+\gamma)^0 \, b}{2\pi}\, \e^{(a_1x+a_2y)}\, K_0(a r)\, .
\end{align}
In addition, we use the ansatz for $F$
\begin{align}
F=F^0+\e^{(a_1 x+a_2 y)} \Phi
\end{align}
and obtain from (\ref{F-pde}) the Helmholtz equation for $\Phi$
\begin{align}
\label{Phi-pde}
[\Delta-\kappa^2]\,\Phi&= -(\mu+\gamma)^0\, b\, \delta(x)\delta(y)\,.
\end{align}
The solution of (\ref{Phi-pde}) is given by
\begin{align}
\Phi=\frac{(\mu+\gamma)^0\, b}{2\pi}\, K_0(\kappa r)\, .
\end{align}
Eventually, the stress function $F$ is calculated as
\begin{align}
\label{F}
F=-\frac{(\mu+\gamma)^0\, b}{2\pi}\, \e^{(a_1 x+a_2 y)}\, \big[K_0( ar)-K_0(\kappa r)\big]\,.
\end{align}

\begin{figure}[pt]\unitlength1cm
\centerline{
\epsfig{figure=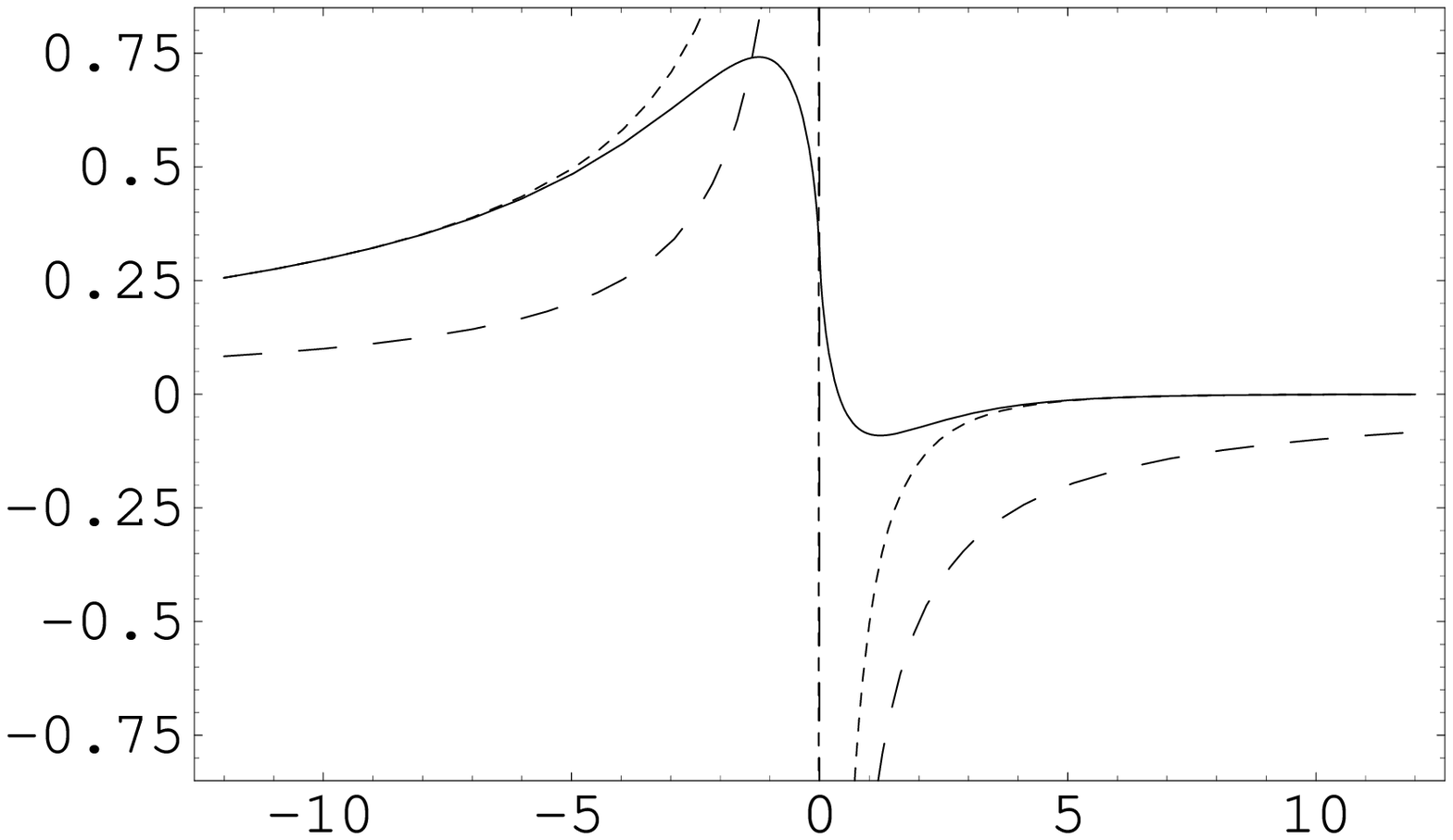,width=7.0cm}
\put(-7.5,2.0){$\beta_{zx}$}
\put(-6.8,-0.3){$\text{(a)}$}
\hspace*{0.2cm}
\put(0,-0.3){$\text{(b)}$}
\epsfig{figure=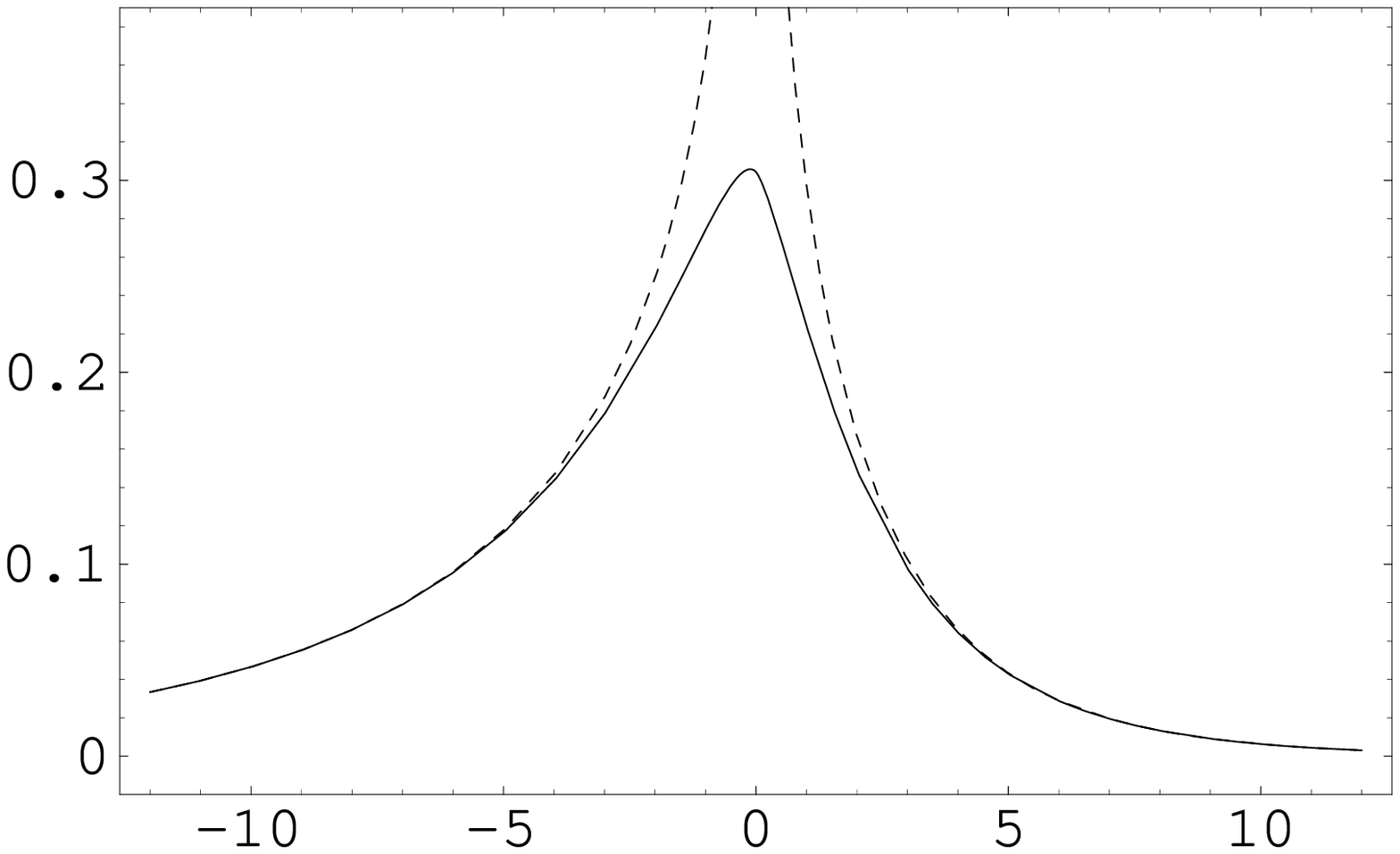,width=6.8cm}
}
\vspace*{0.2cm}
\centerline{
\epsfig{figure=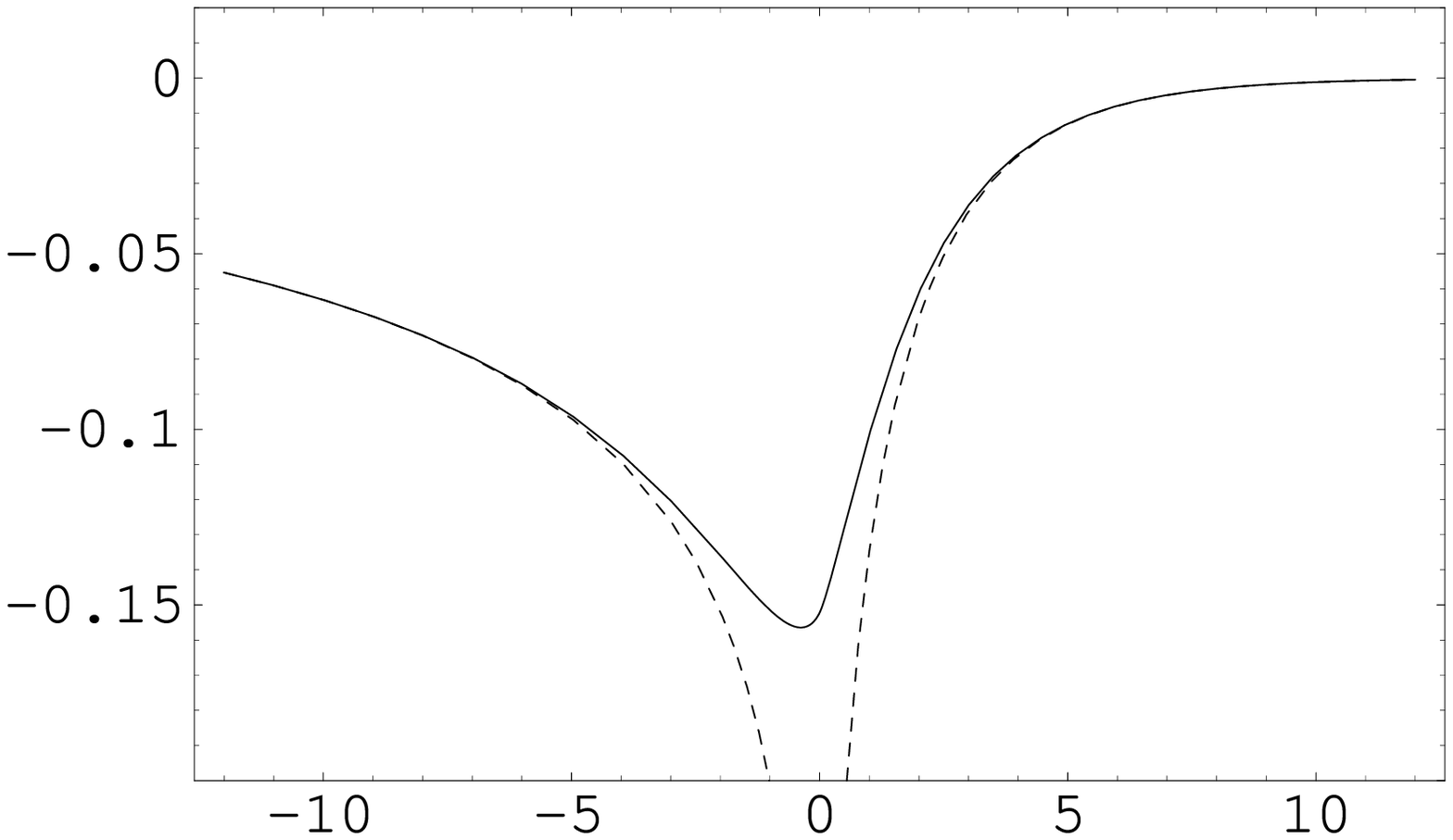,width=7.0cm}
\put(-3.3,-0.3){$ y/\ell$}
\put(-7.5,3.2){$ \beta_{zy}$}
\put(-6.8,-0.3){$\text{(c)}$}
\hspace*{0.2cm}
\put(3.6,-0.3){$x/\ell$}
\put(0,-0.3){$\text{(d)}$}
\epsfig{figure=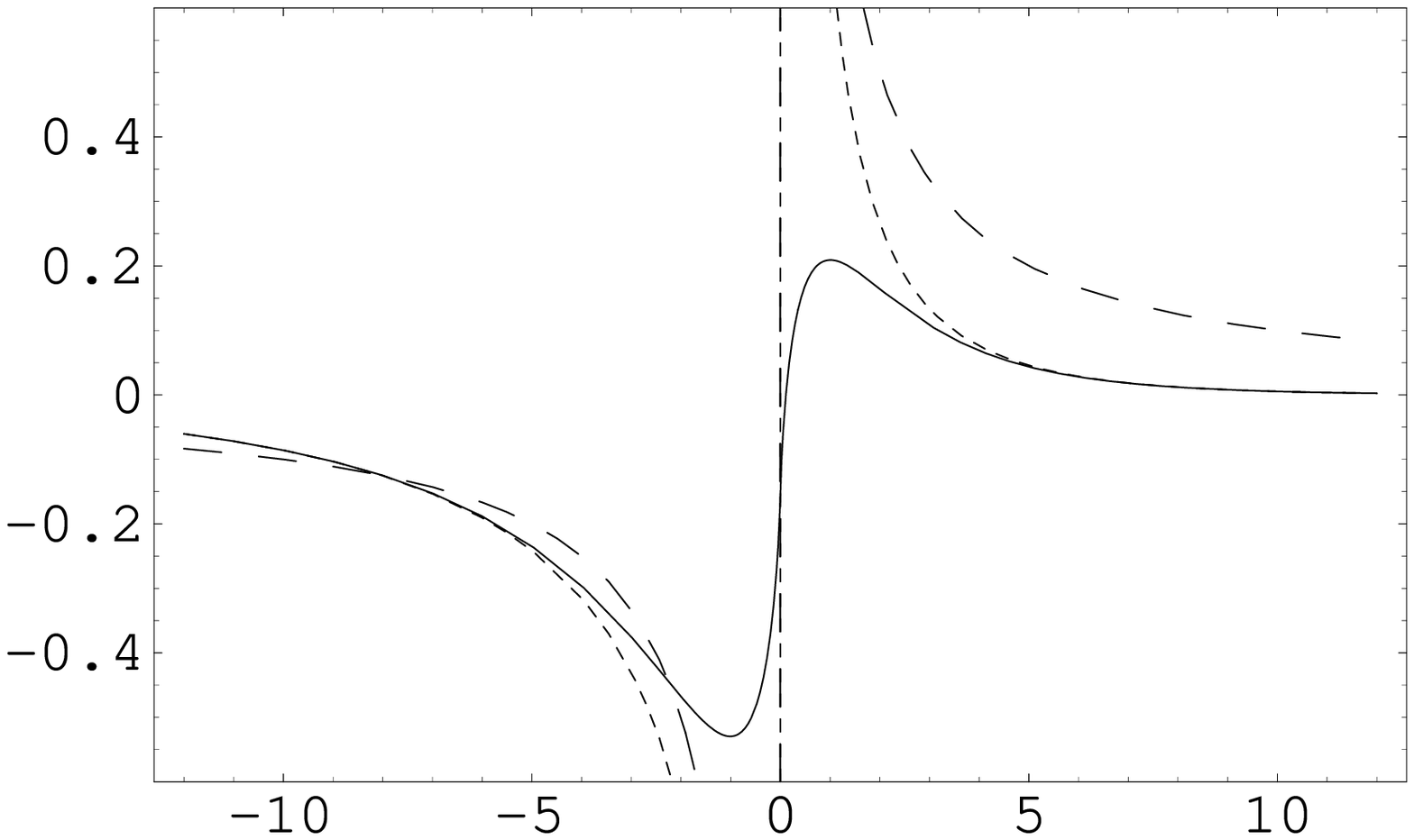,width=7.0cm}
}
\caption{Elastic distortion of a screw dislocation in FGM
using gauge theory (solid line)
and elasticity (small dashed line)
are given in units of $b/[2\pi]$ and with $a_1=0.1/\ell$ and
$a_2=0.2/\ell$:
(a) $\beta_{zx}(0,y)$,
(b) $\beta_{zx}(x,0)$,
(c) $\beta_{zy}(0,y)$,
(d) $\beta_{zy}(x,0)$.
The dashed curves represent the elastic distortion in classical elasticity of a
homogeneous medium.}
\label{fig:strain}
\end{figure}
\begin{figure}[pt!!!]\unitlength1cm
\centerline{
\epsfig{figure=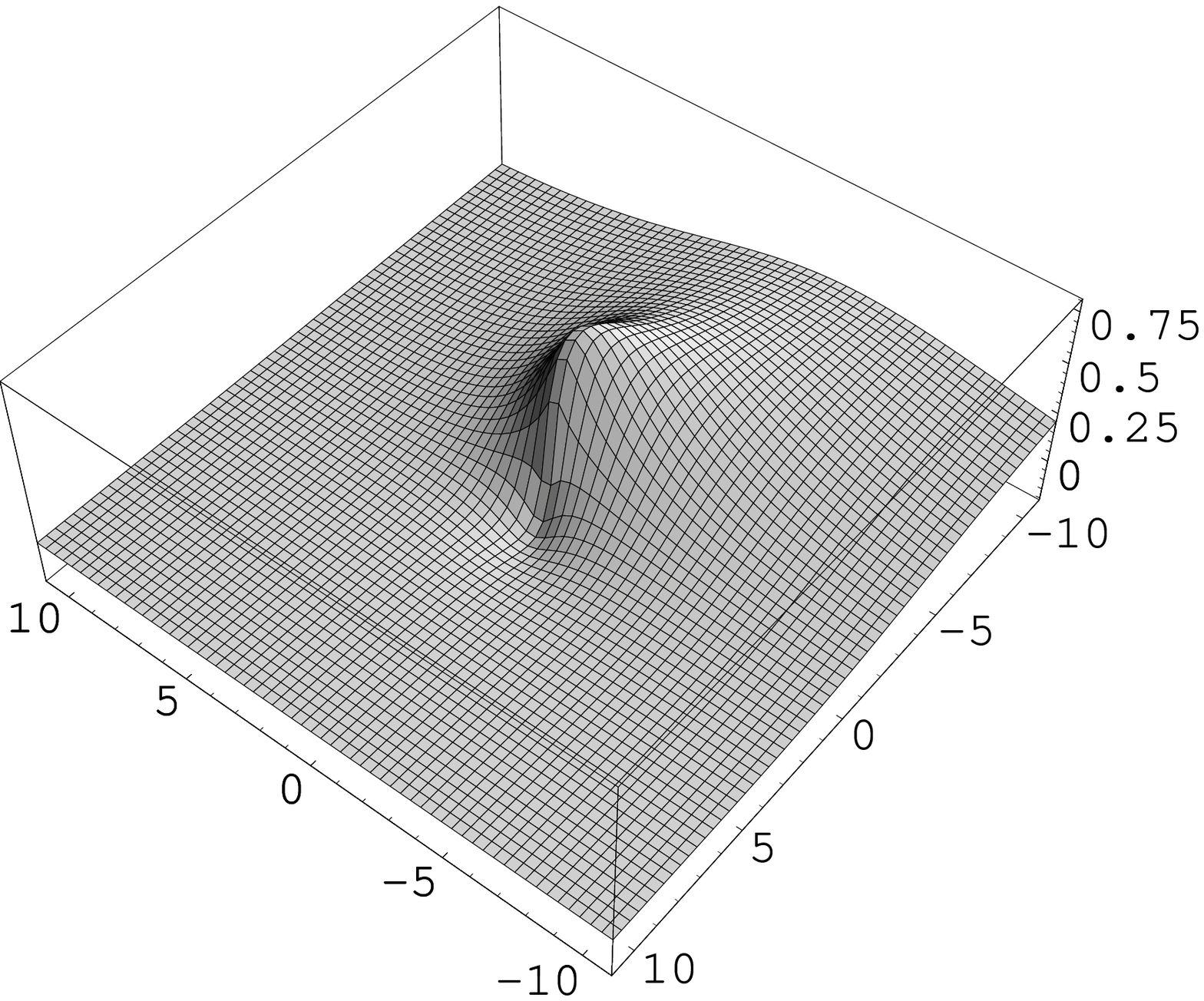,width=7.0cm}
\put(-1.5,1.0){$\frac{y}{\ell}$}
\put(-6.5,1.0){$\frac{x}{\ell}$}
\put(-6.2,-0.3){$\text{(a)}$}
\hspace*{0.4cm}
\put(0,-0.3){$\text{(b)}$}
\put(5.5,1.0){$\frac{y}{\ell}$}
\put(0.5,1.0){$\frac{x}{\ell}$}
\epsfig{figure=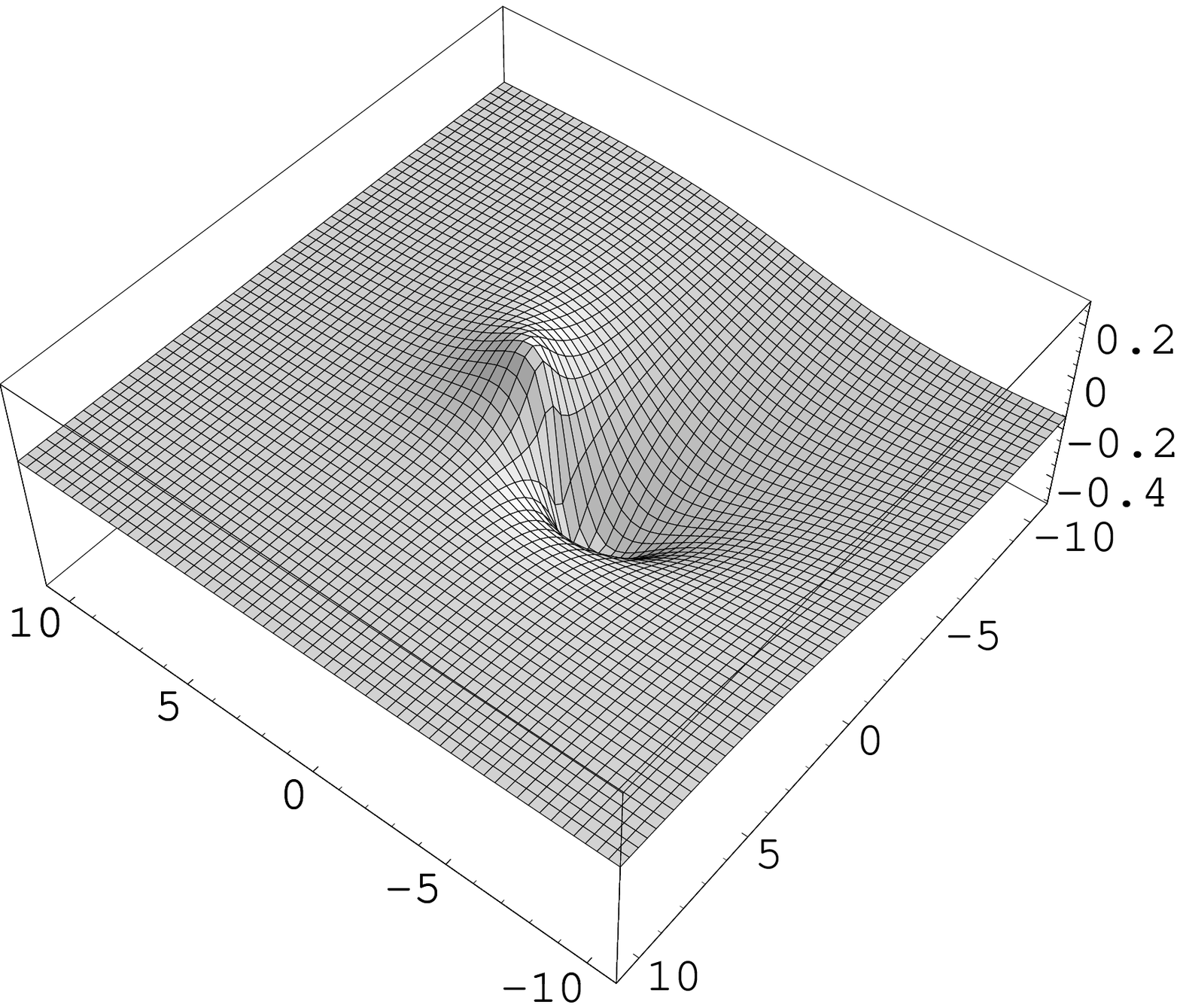,width=7.0cm}
}
\caption{Three-dimensional plots
 of the elastic distortion of a screw dislocation in FGMs in units of 
 $b/[2\pi]$ ($a_1=0.1/\ell$ and $a_2=0.2/\ell$): 
(a) $\beta_{zx}$ and (b) $\beta_{zy}$.}
\label{fig:B3D}
\end{figure}

Substituting (\ref{F}) into (\ref{Bzx-F}) and (\ref{Bzy-F}), we arrive at the expressions 
of the elastic distortions as
\begin{align}
\label{Bzx}
\beta_{zx}&=-\frac{b}{2\pi}\, 
\e^{-(a_1 x+a_2 y)} \Big(\frac{y}{r}\big[a K_1(a r)-\kappa K_1(\kappa r)\big]
-a_2 \big[K_0( ar)-K_0(\kappa r)\big]\Big)\, ,\\
\label{Bzy}
\beta_{zy}&=\frac{b}{2\pi}\, 
\e^{-(a_1 x+a_2 y)} \Big(\frac{x}{r}\big[a K_1(a r)-\kappa K_1(\kappa r)\big]
-a_1 \big[K_0( ar)-K_0(\kappa r)\big]\Big)\, .
\end{align}
The elastic distortions $\beta_{zx}$ and $\beta_{zy}$ are plotted 
in Figs.~(\ref{fig:strain}) and (\ref{fig:B3D}).
First of all, it can be seen that the elastic distortions~(\ref{Bzx}) and
(\ref{Bzy}) are nonsingular unlike the result for FGMs using elasticity theory
and that they are given in terms of two inverse length scales $a$ and $\kappa$.
The component $\beta_{zx}(0,y)$ has extremum values near the dislocation line
at $y=0$ and it has a finite value at $y=0$ depending on the material moduli.
Unlike homogeneous media, it does not satisfy:
$\beta_{zx}(0,y)\neq-\beta_{zx}(0,-y)$.
Also $\beta_{zx}(x,0)$ possesses a maximum at $x=0$ and 
is asymmetric with respect to the plane $x=0$: 
$\beta_{zx}(x,0)\neq\beta_{zx}(-x,0)$. 
Analogously, $\beta_{zy}(x,0)$ has extremum values near the dislocation line 
at $x=0$ and it has a finite value at $x=0$. 
It does not fulfill,
$\beta_{zy}(x,0)\neq-\beta_{zy}(-x,0)$, as for homogeneous materials.
Evidently, $\beta_{zx}(y,0)$ has a minimum at $y=0$
and is asymmetric with respect to the plane $y=0$: 
$\beta_{zy}(0,y)\neq\beta_{zy}(0,-y)$.
For FGMs the far-field of the 
curve of the elastic distortion $\beta_{zx}$ is for negative $y$ 
higher than the $-1/y$ behavior known from homogeneous materials (see Fig.~\ref{fig:strain}a). Due to the gradation of the material the elastic distortion 
in FGMs decays faster than in homogeneous media.

\begin{figure}[pt]\unitlength1cm
\centerline{
\epsfig{figure=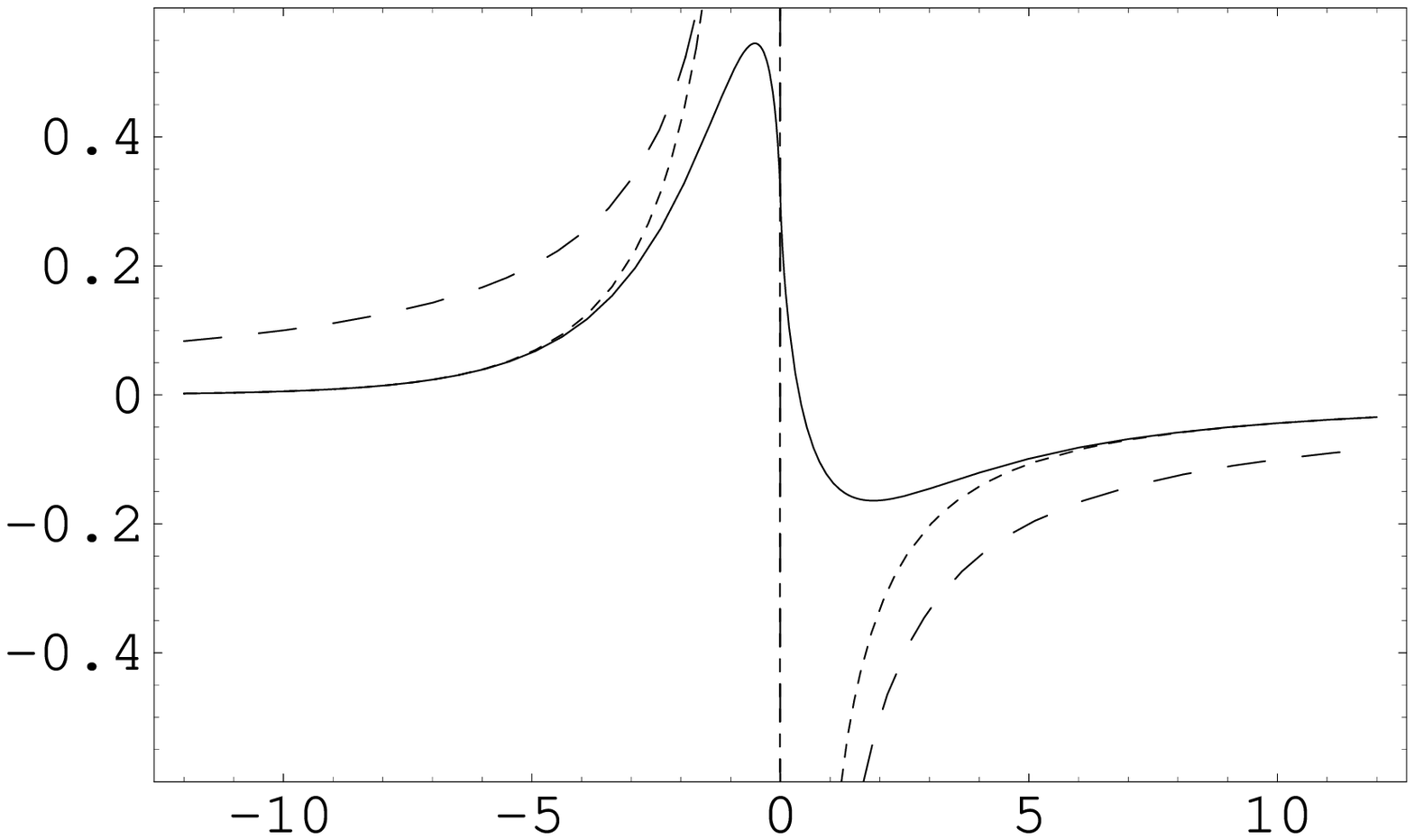,width=7.0cm}
\put(-7.5,2.0){$\sigma_{zx}$}
\put(-6.8,-0.3){$\text{(a)}$}
\hspace*{0.2cm}
\put(0,-0.3){$\text{(b)}$}
\epsfig{figure=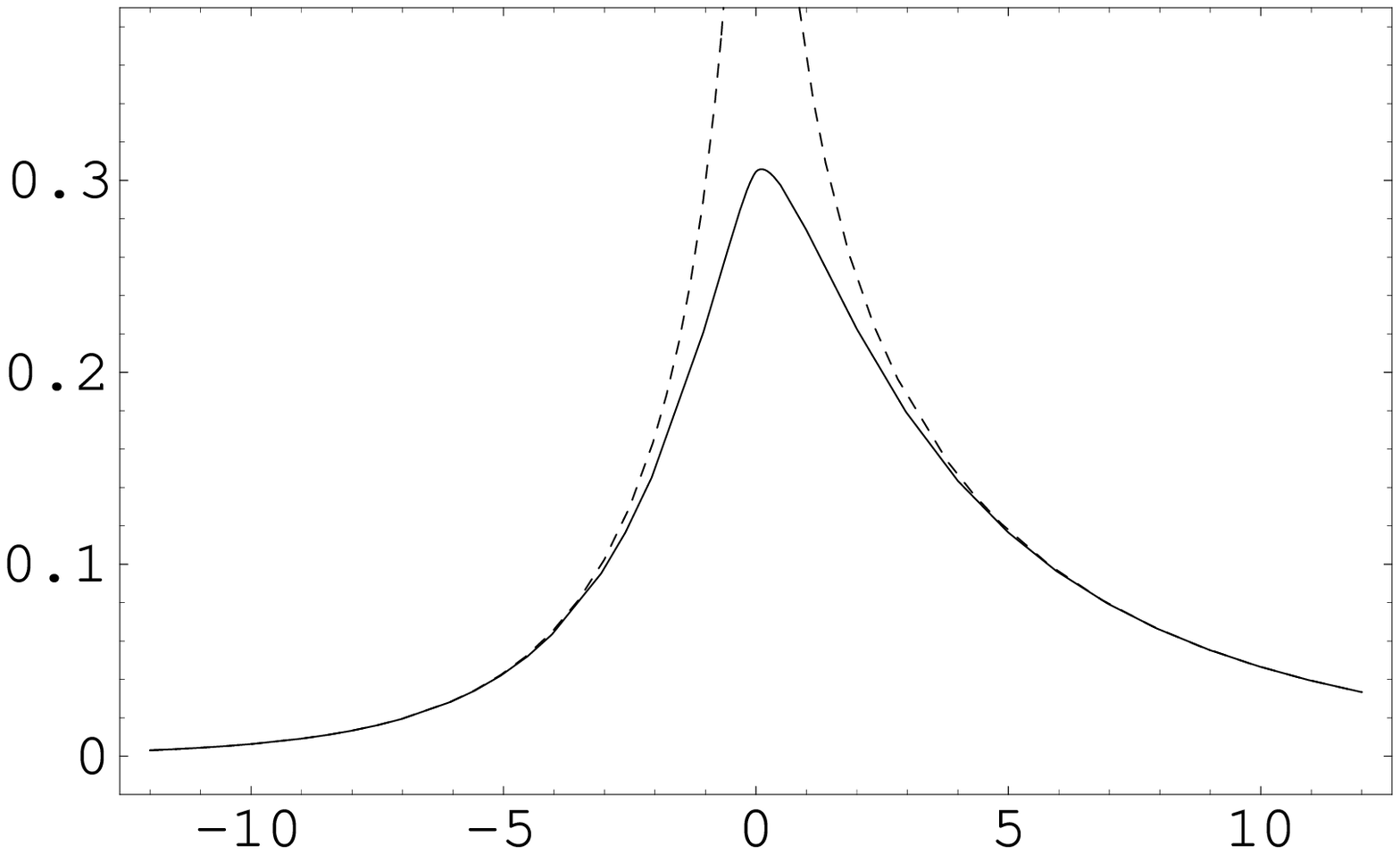,width=6.8cm}
}
\vspace*{0.2cm}
\centerline{
\epsfig{figure=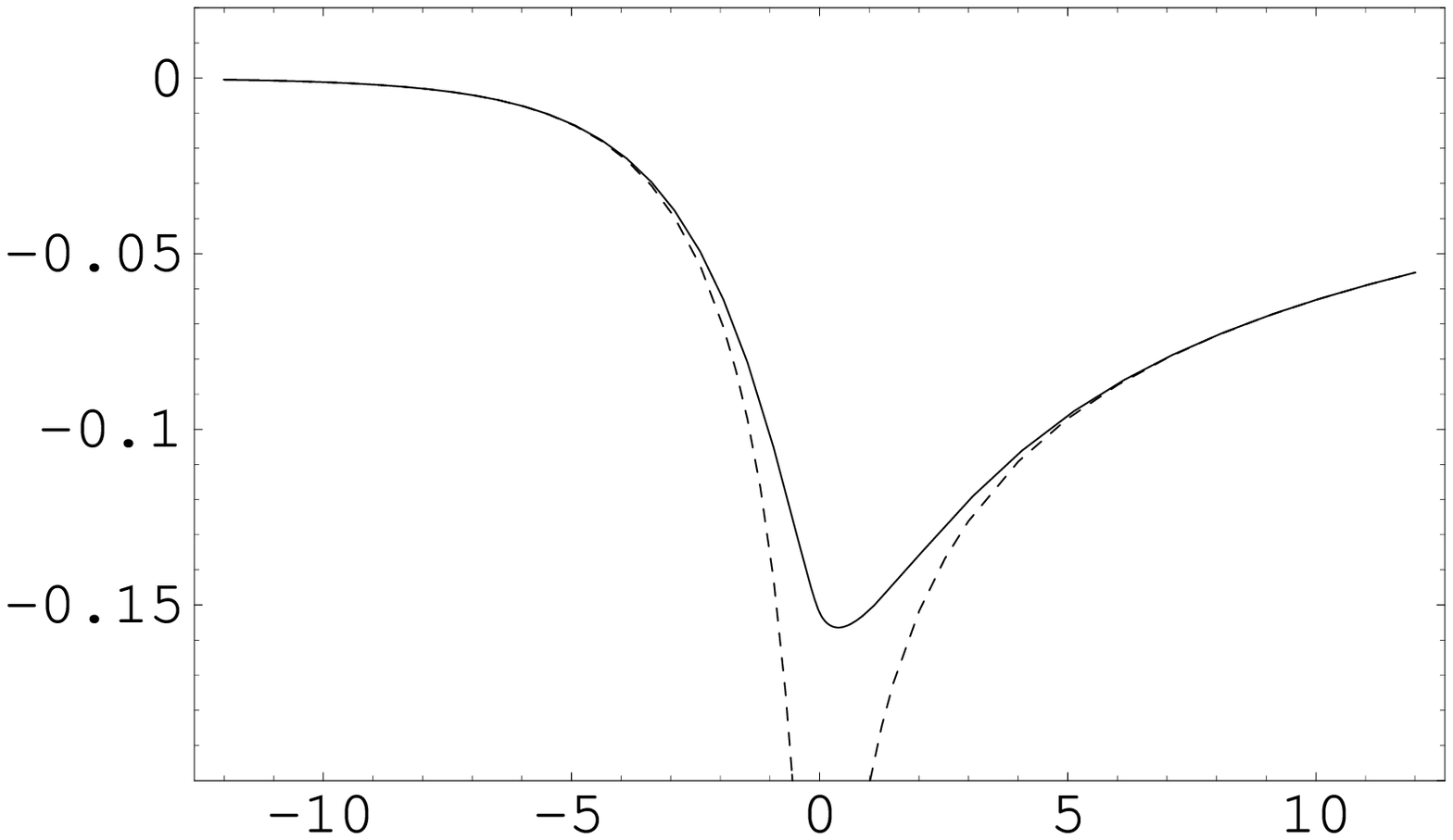,width=7.2cm}
\put(-3.4,-0.3){$y/\ell$}
\put(-7.5,3.2){$ \sigma_{zy}$}
\put(-6.8,-0.3){$\text{(c)}$}
\hspace*{0.2cm}
\put(3.6,-0.3){$x/\ell$}
\put(0,-0.3){$\text{(d)}$}
\epsfig{figure=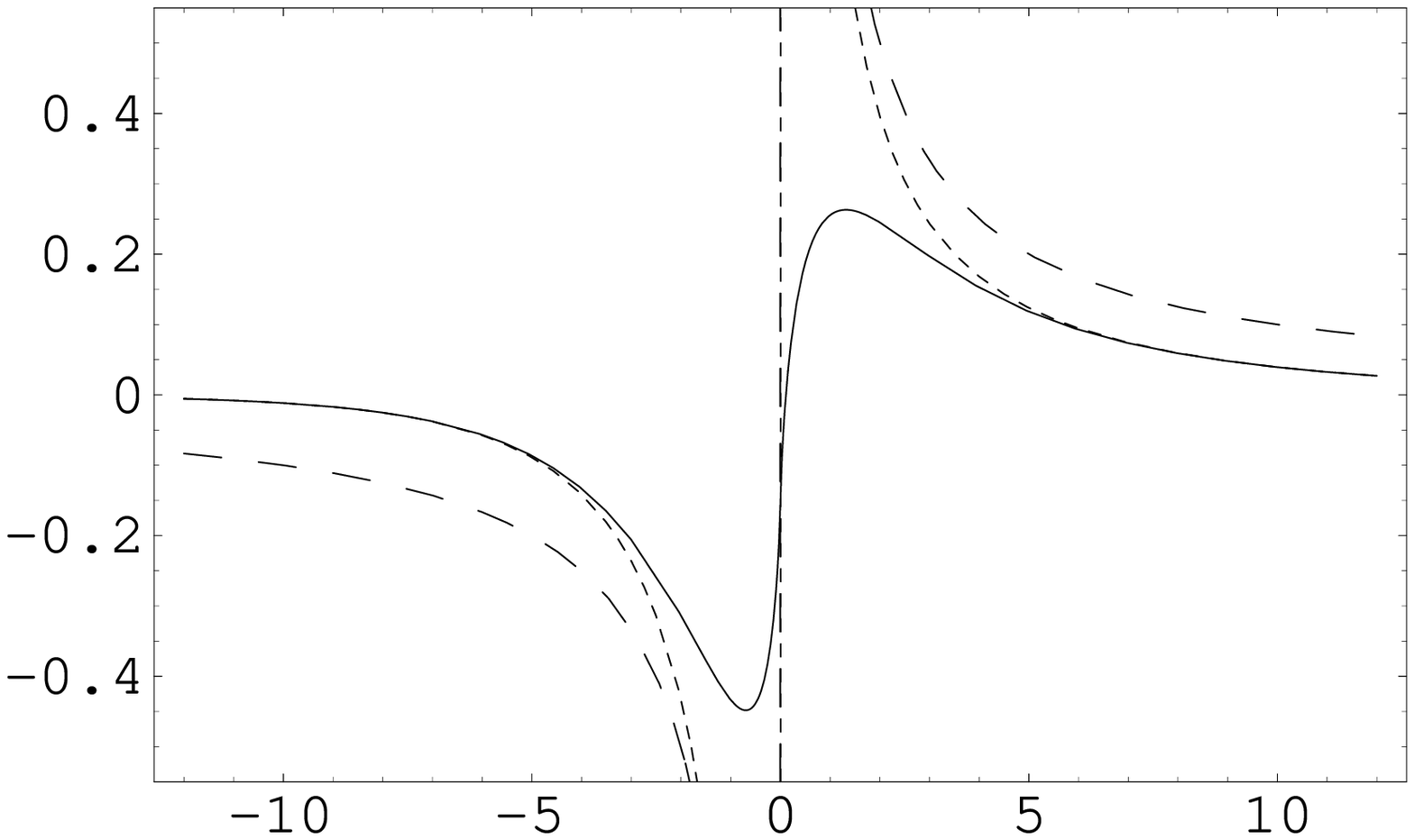,width=7.0cm}
}
\caption{Stress of a screw dislocation in FGM
using gauge theory (solid line)
and elasticity (small dashed line)
are given in units of $(\mu+\gamma)_0\, b/[2\pi]$ and with $a_1=0.1/\ell$
and $a_2=0.2/\ell$:
(a) $\sigma_{zx}(0,y)$,
(b) $\sigma_{zx}(x,0)$,
(c) $\sigma_{zy}(0,y)$,
(d) $\sigma_{zy}(x,0)$.
The dashed curves represent the stress in classical elasticity of a
homogeneous medium.}
\label{fig:stress}
\end{figure}

\begin{figure}[pt!]\unitlength1cm
\centerline{
\epsfig{figure=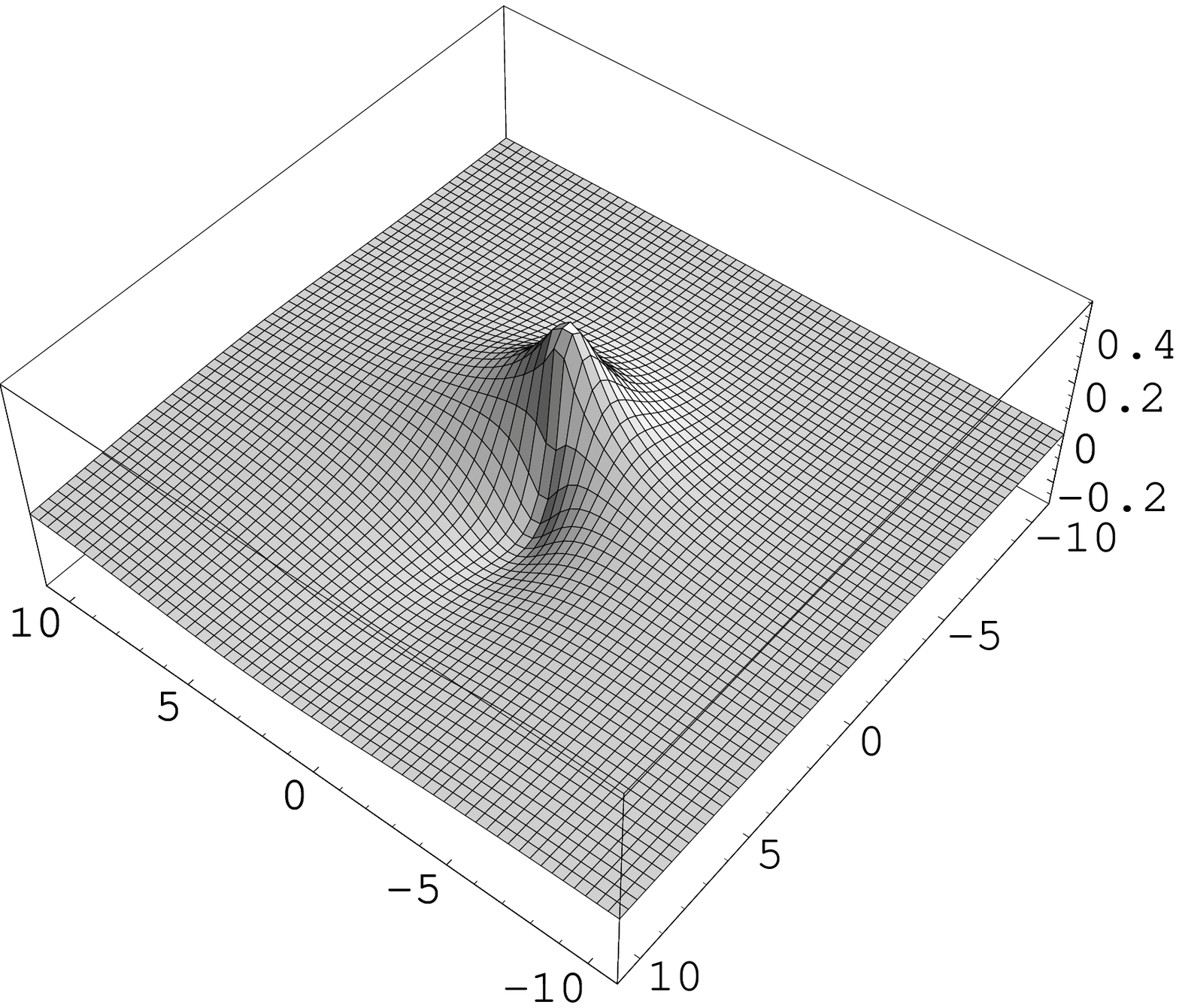,width=7.0cm}
\put(-1.5,1.0){$\frac{y}{\ell}$}
\put(-6.5,1.0){$\frac{x}{\ell}$}
\put(-6.2,-0.3){$\text{(a)}$}
\hspace*{0.4cm}
\put(0,-0.3){$\text{(b)}$}
\put(5.5,1.0){$\frac{y}{\ell}$}
\put(0.5,1.0){$\frac{x}{\ell}$}
\epsfig{figure=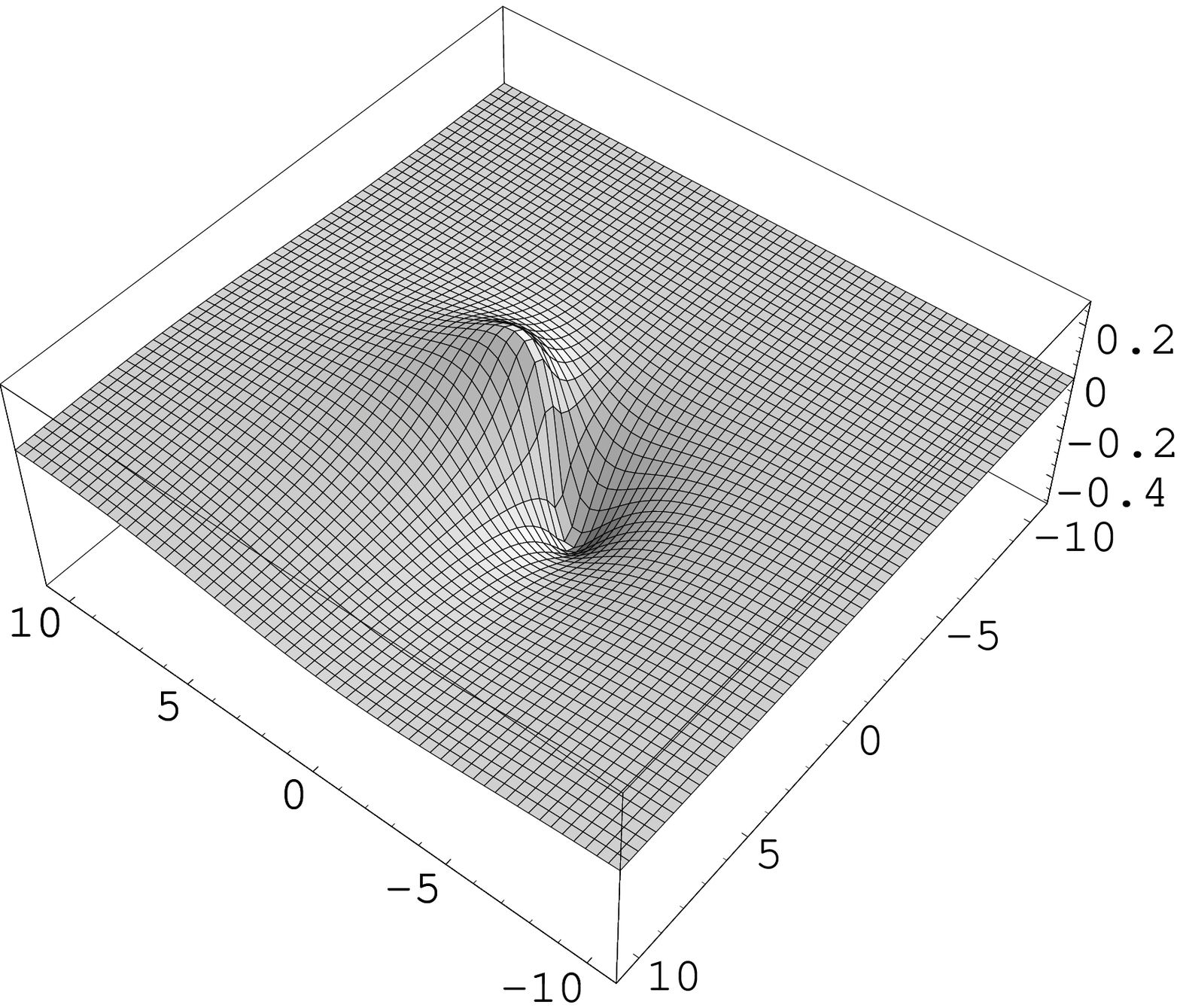,width=7.0cm}
}
\caption{Three-dimensional plots
 of the stress of a screw dislocation in FGMs in units of 
 $(\mu+\gamma)_0\, b/[2\pi]$ ($a_1=0.1/\ell$, $a_2=0.2/\ell$): 
(a) $\sigma_{zx}$ and (b) $\sigma_{zy}$.}
\label{fig:T3D}
\end{figure}

Using (\ref{CE-1}), the stresses can be derived to be
\begin{align}
\label{Tzx}
\sigma_{zx}&=-\frac{(\mu+\gamma)^0\, b}{2\pi}\, 
\e^{(a_1 x+a_2 y)} \Big(\frac{y}{r}\big[a K_1(a r)-\kappa K_1(\kappa r)\big]
-a_2 \big[K_0( ar)-K_0(\kappa r)\big]\Big)\, ,\\
\label{Txz}
\sigma_{xz}&=-\frac{(\mu-\gamma)^0\, b}{2\pi}\, 
\e^{(a_1 x+a_2 y)} \Big(\frac{y}{r}\big[a K_1(a r)-\kappa K_1(\kappa r)\big]
-a_2 \big[K_0( ar)-K_0(\kappa r)\big]\Big)\, ,\\
\label{Tzy}
\sigma_{zy}&=\frac{(\mu+\gamma)^0\,b}{2\pi}\, 
\e^{(a_1 x+a_2 y)} \Big(\frac{x}{r}\big[a K_1(a r)-\kappa K_1(\kappa r)\big]
-a_1 \big[K_0( ar)-K_0(\kappa r)\big]\Big)\, ,\\
\label{Tyz}
\sigma_{yz}&=\frac{(\mu-\gamma)^0\,b}{2\pi}\, 
\e^{(a_1 x+a_2 y)} \Big(\frac{x}{r}\big[a K_1(a r)-\kappa K_1(\kappa r)\big]
-a_1 \big[K_0( ar)-K_0(\kappa r)\big]\Big)\, .
\end{align}
The stresses $\sigma_{zx}$ and $\sigma_{zy}$ are plotted 
in Figs.~(\ref{fig:stress}) and (\ref{fig:T3D}).
It is important to note that the stresses~(\ref{Tzx})--(\ref{Tyz}) 
do not possess any singularity.   
Due to the gradation, the components~(\ref{Tzx})--(\ref{Tyz}) have extremum
values near the dislocation line and a finite value at the dislocation line.
For FGMs the far-field of the 
curves of the stresses are less  than the $1/r$ behavior known from homogeneous materials (see Fig.~\ref{fig:stress}a and \ref{fig:stress}d).

We note that the expressions (\ref{Bzx})--(\ref{Tyz}) coincide 
for $a_1\rightarrow 0$ and $a_2\rightarrow 0$ with the gauge-theoretical 
result obtained by~\citet{LA09} for homogeneous materials and for 
$a_1\rightarrow 0$ and $\gamma\rightarrow 0$ with the result given by~\citet{Lazar07} using the strain gradient elasticity formulation for FGMs.

\section{Conclusion}
In this paper, we have investigated a straight screw dislocation in
FGMs using the gauge theory of dislocations. 
In this framework, we have derived the field equations 
of dislocations for general nonhomogeneous media 
where the constitutive moduli depend on $\Bx$. 
Because of the gradation, the field equations contain gradients
of the constitutive moduli.
Subsequent we have specialized the field equations 
to exponentially graded materials.
We found a new effective inverse length scale $\kappa$ characteristic for
the gauge-theoretical anti-plane problem of FGMs.
We have calculated the elastic distortion, the force stress, the torsion and
the pseudomoment stress fields produced by a 
screw dislocation in FGMs in the framework of dislocation gauge theory.
All the analytical solutions are given in terms of the effective length scale
$1/\kappa$ as well as the gradation length scale $1/a$.
We found `perturbed' Helmholtz equations as governing partial 
differential equations which we solved.
We found exact analytical gauge-theoretical solutions which can be used 
by material scientists and design and manufacturing engineers.
One hope is that these gauge solutions will provide new physical insight to possible
improvements of designing FGMs.
In the limit $\Ba\rightarrow 0$, we recover well-known results
of a screw dislocation in the gauge theory for homogeneous media.

\subsection*{Acknowledgement}
The author has been supported by an Emmy-Noether grant of the 
Deutsche Forschungsgemeinschaft (Grant No. La1974/1-3)
and a Heisenberg fellowship (Grant No. La1974/2-1).

\end{document}